\documentclass[a4paper,11pt]{article}
\usepackage{jheppub}

\usepackage{lineno}
\usepackage{float}
\usepackage{graphicx}
\usepackage{dcolumn}
\usepackage{bm}
\usepackage[dvipsnames]{xcolor}
\usepackage{hyperref}
\usepackage{lineno}
\usepackage[table,x11names]{xcolor}
\usepackage{longtable} 
\usepackage{tabularx}
\usepackage{multirow}
\usepackage[normalem]{ulem}
\usepackage{booktabs}
\usepackage{siunitx}

\title{Multi-Component Dark Matter as a Solution to the Galactic Center GeV Excess}

\author[a,b,c,d]{Farinaldo S. Queiroz}
\author[e]{Clarissa Siqueira}
\author[f]{Carlos E. Yaguna}

\affiliation[a]{Departamento de Física, Universidade Federal do Rio Grande do Norte, 59078-970, Natal, RN, Brasil}
\affiliation[b]{International Institute of Physics, Universidade Federal do Rio Grande do Norte,
Campus Universitario, Lagoa Nova, Natal-RN 59078-970, Brazil}
\affiliation[c]{Millennium Institute for Subatomic Physics at the High-Energy Frontier (SAPHIR) of ANID, Fern\'andez Concha 700, Santiago, Chile}
\affiliation[d]{Departamento de F\'isica, Facultad de Ciencias, Universidad de La Serena,
Avenida Cisternas 1200, La Serena, Chile}
\affiliation[e]{Observatório Nacional, Rio de Janeiro - RJ, 20921-400, Brasil}
\affiliation[f]{Escuela de Física, Universidad Pedagógica y Tecnológica de Colombia,
Avenida Central del Norte \# 39-115, Tunja, Colombia}

\emailAdd{csiqueira@on.br}

\abstract{\noindent
The Galactic Center Excess (GCE) is a compelling signature of dark matter annihilation, but its spectral morphology is difficult to reconcile with the traditional paradigm of a single particle species. In this work, we perform a systematic investigation of multi-component dark matter sectors, exploring scenarios with two ($N=2$) and three ($N=3$) distinct particle species while considering both exclusive and mixed annihilation channels. Using the Akaike Information Criterion (AIC) to rigorously penalize model complexity, we find that the GCE data statistically favors an $N=2$ scenario where each dark matter component annihilates exclusively into a single final state. Our results reveal that the preferred solutions naturally follow a light-plus-heavy mass hierarchy, and that specific final states such as $t\bar{t}$, $ZZ$, and $hh$, which are individually unable to explain the excess are effectively ``resurrected'' by the improved morphological fit provided by the multi-component framework. Furthermore, we show that these scenarios may mitigate the tension with current constraints, reaching compatibility within existing uncertainties. Our results suggest that the GCE may be the first evidence of a diverse dark sector, favoring a multi-scale solution over the minimal WIMP paradigm.}

\begin{document}
\maketitle
\flushbottom

\section{Introduction}
\label{sec:intro}

Analyses of gamma-ray data collected by the Fermi Large Area Telescope (Fermi-LAT) have consistently identified a spatially extended excess of emission radiating from the inner Galaxy \cite{Goodenough:2009gk,Hooper:2010mq,Hooper:2011ti,Abazajian:2012pn}. This feature, widely referred to as the Galactic Center Excess (GCE), is characterized by an intensity exceeding the predictions of standard Galactic diffuse emission models and of known point-source catalogs \cite{Muru:2025vpz,Fermi-LAT:2016zaq,Fermi-LAT:2022byn}. The excess is approximately spherically symmetric, extending roughly $10^\circ$ from the Galactic Center, with a spatial morphology that follows the square of a generalized Navarro-Frenk-White (gNFW) density profile \cite{Navarro:1995iw, Navarro:1996gj} with an inner slope $\gamma\approx 1.2$. Spectrally, the emission exhibits a distinct \emph{bump} that peaks between $1$ and $5$ GeV and falls off rapidly at higher energies \cite{Vitale:2011zz,Calore:2014nla,Daylan:2014rsa}.

Two leading hypotheses have emerged to explain the origin of the GCE \cite{Murgia:2020dzu, Slatyer:2021qgc,Kong:2025ccv}. The first attributes the emission to a population of unresolved astrophysical point sources, most notably Millisecond Pulsars (MSPs), which are known to emit gamma rays with a spectral shape similar to the observed excess \cite{Abazajian:2010zy,Bartels:2015aea,Lee:2015fea,Macias:2016nev,Bartels:2017vsx,Lei:2025jsu}. While there is significant support for the MSP hypothesis, particularly from statistical analyses of photon clustering, the debate regarding their luminosity function and spatial distribution remains active \cite{Amerio:2024qor,Holst:2024fvb,Zhong:2019ycb}. Alternatively, the GCE has been widely interpreted as a signal of dark matter (DM) particle annihilation. This scenario is particularly compelling because the observed spatial morphology matches the predictions for a dark matter halo, and the spectral peak is remarkably well-described by Weakly Interacting Massive Particles (WIMPs) in the $30$ to $50$ GeV mass range annihilating into $b\bar b$ quarks \cite{Ruiz-Alvarez:2012nvg,Agrawal:2014una, DiMauro:2021qcf}. In this work, we focus exclusively on the dark matter interpretation, investigating whether the detailed features of the signal actively favor a dark sector that is more sophisticated than the standard single-component WIMP \cite{Calore:2014nla,Leane:2019xiy,Cholis:2021rpp,Kong:2025ccv}.

From a theoretical perspective, there is no fundamental reason to assume the dark sector is populated by a single particle species \cite{Zurek:2008qg,Profumo:2009tb}. The visible Standard Model is rich and complex, containing multiple matter fields with distinct masses and interactions; it is therefore natural to consider that the dark sector may exhibit a similar structure. Multicomponent dark matter scenarios—such as those arising from $Z_N$ symmetries—represent a natural generalization of the minimal single-particle hypothesis while remaining firmly rooted within the WIMP framework \cite{Belanger:2020hyh,Yaguna:2021vhb,Yaguna:2021rds,Belanger:2022esk}. In the context of indirect detection, these scenarios offer distinctive phenomenological signatures that single-component models cannot reproduce. Specifically, the combined emission from multiple mass eigenstates allows for composite spectral shapes that can simultaneously accommodate features at different energy scales, overcoming the kinematic rigidity of single-particle annihilation. Given the high statistical significance and spectral precision of the Fermi-LAT observations, the GCE provides an ideal laboratory to confront the predictions of these multicomponent dark matter scenarios with data.

In this work, we perform a comprehensive phenomenological analysis of the GCE through the lens of multicomponent dark matter. To ensure the highest fidelity in our fits and mitigate systematic uncertainties from Galactic diffuse emission, we restrict our analysis to the cleaner Southern Sky and utilize a recent, robust background covariance matrix \cite{Cholis:2021rpp}. We systematically investigate scenarios containing two and three distinct dark matter particle species, exploring models where each component annihilates exclusively into a single Standard Model final state, as well as more general scenarios permitting mixed annihilation channels. Crucially, by employing the Akaike Information Criterion (AIC) to strictly balance fit quality against model complexity, we demonstrate a strong statistical preference for an exclusive two-component dark matter framework over the standard single-particle hypothesis. The remainder of this paper is organized as follows: Section II outlines the theoretical formulation of the multicomponent dark matter framework. Section III describes the data selection and background modeling used in our analysis. Section IV presents our main results, detailing the best-fit regions, the statistical supremacy of the two-component scenario, and the penalties associated with over-complex models. Finally, we discuss the physical implications of our findings in Section V and offer our conclusions in Section IV.




\section{Theoretical framework}
\label{sec:theory}

We consider a multicomponent dark sector composed of $N$ distinct, stable, and electrically neutral particle species, denoted by $\chi_i$ (where $i = 1, \dots, N$). Each component constitutes a unique physical state characterized by its specific mass, $m_{\chi, i}$, and a distinct set of couplings to Standard Model particles, as well as potential interactions with other dark sector states. This framework naturally encompasses the standard single-component WIMP hypothesis widely analyzed in the literature, which is recovered in the limit $N=1$.

In this framework, the total dark matter abundance observed in the universe today is the sum of the individual relic densities of all stable components \cite{Planck:2018vyg}:
\begin{equation}
    \Omega_{\text{DM}}h^2 = \sum_{i=1}^{N} \Omega_i h^2 \approx 0.12,
\end{equation}
where $\Omega_{\text{DM}}h^2$ represents the total cold dark matter density measured by \textit{Planck}. We define the relative abundance, or ``dark matter fraction'', for each species $i$ as $\xi_i$:
\begin{equation}
    \xi_i \equiv \frac{\Omega_i}{\Omega_{\text{DM}}}.
\end{equation}
By definition, these fractions must satisfy the normalization condition $\sum_{i=1}^N \xi_i = 1$. We assume that the spatial distribution of each dark matter component follows the standard total dark matter density profile of the Galaxy, $\rho_{\text{tot}}(r)$, scaled by its cosmological abundance. Thus, the local density of species $i$ is given by $\rho_i(r) = \xi_i \rho_{\text{tot}}(r)$.

\subsection{Gamma-Ray Flux and Effective Cross-Section}

The differential flux of gamma rays produced by dark matter annihilation is proportional to the square of the dark matter density and the strength of the annihilation processes. In our multicomponent framework, the total differential flux is the sum of the contributions from each species:
\begin{equation}
    \frac{d\phi_\gamma}{dE} = \frac{1}{4\pi} \sum_{i=1}^{N} \frac{\langle \sigma v \rangle_i}{2 m_{\chi, i}^2} \int_{\Delta \Omega} d\Omega \int_{\text{l.o.s}} \rho_i^2(r) \, dl \times \frac{dN_{\gamma, i}}{dE},
    \label{eq:flux}
\end{equation}
where $\langle \sigma v \rangle_i$ is the total thermally averaged annihilation cross-section for species $i$, and $dN_{\gamma, i}/dE$ is the photon spectrum produced per annihilation event.

Substituting the density relation $\rho_i(r) = \xi_i \rho_{\text{tot}}(r)$, we can factor out the astrophysical J-factor, $J(\Delta \Omega) = \int \int \rho_{\text{tot}}^2 dl d\Omega$, which is common to all components (assuming identical spatial profiles). The expression simplifies to:
\begin{equation}
    \frac{d\phi_\gamma}{dE} = \frac{J(\Delta \Omega)}{4\pi} \sum_{i=1}^{N} \left( \xi_i^2 \langle \sigma v \rangle_i \right) \frac{1}{2 m_{\chi, i}^2} \frac{dN_{\gamma, i}}{dE}.
\end{equation}
It is crucial to note that the gamma-ray flux depends on the product of the squared fraction and the cross-section, $\xi_i^2 \langle \sigma v \rangle_i$. Consequently, indirect detection experiments alone cannot disentangle the intrinsic interaction strength from the cosmological abundance. In our analysis, we therefore report results in terms of the \textbf{effective cross-section} for each component:
\begin{equation}
    \langle \sigma v \rangle_{\text{eff}, i} \equiv \xi_i^2 \langle \sigma v \rangle_i.
\end{equation}
This quantity represents the experimentally accessible observable, encompassing both the particle physics couplings and the relative density of the species.

\subsection{Model Parameters and Degrees of Freedom}

In this analysis, we treat the number of dark matter components, $N$, as a discrete model choice. For a given $N$, the physics of the dark sector is determined by the mass of each particle, its total effective annihilation strength, and the specific composition of its Standard Model final states. 

For a generic component $\chi_i$, the total photon spectrum $dN_{\gamma, i}/dE$ is given by the weighted sum over all open annihilation channels $f$:
\begin{equation}
    \frac{dN_{\gamma, i}}{dE} = \sum_{f} B_{i,f} \frac{dN_{\gamma}^{f}}{dE},
\end{equation}
where $dN_{\gamma}^{f}/dE$ is the spectrum for the specific final state $f$ (e.g., $b\bar{b}, \tau^+\tau^-$) and $B_{i,f}$ is the branching ratio into that channel, subject to the normalization condition $\sum_f B_{i,f} = 1$.

Consequently, the set of free parameters for each component $i$ is:
\begin{itemize}
    \item The dark matter mass, $m_{\chi, i}$.
    \item The total effective annihilation cross-section, $\langle \sigma v \rangle_{\text{eff}, i} = \xi_i^2 \langle \sigma v \rangle_i$.
    \item The set of independent branching ratios $\{ B_{i,f} \}$ into the relevant Standard Model final states.
\end{itemize}

We investigate two classes of models. First, we consider simplified benchmarks where each component is assumed to annihilate dominantly into a single final state ($B_{i,f} \approx 1$). While generic WIMP models typically predict a mixture of final states, these exclusive channels serve as highly predictive phenomenological proxies that introduce exactly two degrees of freedom per component. Second, for completeness, we relax this assumption and perform a general analysis where multiple annihilation channels are permitted with free branching ratios. By rigorously tracking these degrees of freedom, we establish the statistical foundation necessary to penalize over-complexity and conduct a fair comparison against the single-component baseline prevalent in the literature.

\section{Methodology}
\label{sec:analysis}

To quantitatively confront our multicomponent dark matter models with the Fermi-LAT observations, we evaluate the goodness-of-fit using a $\chi^2$ test statistic, defined as:
\begin{equation}
    \chi^2 = \sum_{i,j=1}^{14} \left( E_i^2f_i^{\text{GCE}} - E_i^2f^{\text{DM}}_{ik}(\theta_k) \right) \mathcal{C}_{ij}^{-1} \left( E_j^2f_j^{\text{GCE}} - E_j^2f^{\text{DM}}_{jk}(\theta_k) \right),
\end{equation}
where $E_{i,j}$ represents the energy in each bin (spanning 14 energy bins), $f_i^{\text{GCE}}$ is the observed Galactic Center Excess flux measured by Fermi-LAT, and $f^{\text{DM}}_{ik}(\theta_k)$ is the theoretical dark matter flux given in Eq.~\ref{eq:flux}. The vector $\theta_k$ denotes the free parameters of the chosen dark matter model. The matrix $\mathcal{C}_{ij}$ is the covariance matrix, which rigorously accounts for both statistical uncertainties and correlated systematic errors across the energy bins.

We determine the best-fit parameters by minimizing the $\chi^2$ function using the \texttt{NMinimize} routine in Mathematica, yielding $\chi^2_{\text{min}}$. To assess the absolute goodness-of-fit, the $p$-value is computed via $p\text{-value} = 1 - \text{CDF}_{\chi^2}(\chi^2_{\text{min}}, \nu)$, where CDF is the cumulative distribution function for a $\chi^2$ distribution and $\nu$ is the number of degrees of freedom (calculated as the number of data bins minus the number of free model parameters). 

Crucially, to perform a robust statistical comparison between models with different numbers of components, we employ the Akaike Information Criterion (AIC). The AIC balances the goodness-of-fit against the complexity of the model, effectively penalizing the addition of unnecessary parameters:
\begin{equation}
    \text{AIC} = \chi^2_{\text{min}} + 2k,
\end{equation}
where $k$ is the total number of free parameters in the model. We quantify the preference for a multicomponent model over the standard single-component ($N=1$) baseline using the difference $\Delta \text{AIC} = \text{AIC}_{\text{model}} - \text{AIC}_{N=1}$. To evaluate the relative quality of the models, we adopt the widely recognized scale for model selection \cite{Burnham:2002}. Within this framework, a value of $\Delta \text{AIC} < -2$ indicates positive support for the multicomponent model, while values in the range $-7 \le \Delta \text{AIC} \le -4$ represent substantial to strong evidence against the baseline model. Values of $\Delta \text{AIC} < -10$ are considered decisive in favor of the multicomponent scenario.

The covariance matrix $\mathcal{C}_{ij}$ used in this work was derived by Cholis et al. \cite{Cholis:2021rpp}, who conducted a comprehensive analysis of Fermi-LAT data in the central region of our Galaxy. While their analysis covers the full $40^\circ \times 40^\circ$ sky map as well as independent northern and southern hemispheres, we restrict our primary analysis to the Southern Sky data. The choice of this region has profound implications for the robustness of the dark matter interpretation; the Northern Sky is heavily contaminated by complex Galactic diffuse emission, whereas the Southern Sky provides a significantly cleaner environment, thereby minimizing systematic uncertainties \cite{Cholis:2021rpp}.

The theoretical dark matter flux $f^{\text{DM}}_{ik}(\theta_k)$ was computed by treating the dark matter mass ($m_{\chi,i}$), the effective velocity-averaged annihilation cross-section ($\langle \sigma v \rangle_{\text{eff},i}$), and, when applicable, the branching fractions ($B_{i,f}$) as free parameters. To compute the astrophysical $J$-factor, we assume a contracted Navarro-Frenk-White (cNFW) halo profile for the dark matter spatial distribution:
\begin{equation}
    \rho_{\text{cNFW}}(r) = \rho_s \left( \frac{r}{r_s} \right)^{-\gamma_{\text{cNFW}}} \left( 1 + \frac{r}{r_s} \right)^{-(3 - \gamma_{\text{cNFW}})},
\end{equation}
where we adopt an inner slope of $\gamma_{\text{cNFW}} = 1.26$ and a scale radius of $r_s = 20$ kpc. The scale density $\rho_s \approx 0.252$ GeV/cm$^3$ is normalized such that the local dark matter density at the solar position ($r = 8.5$ kpc) is $\rho(8.5 \text{ kpc}) = 0.4$ GeV/cm$^3$.

\section{Results}
\label{sec:results}

\subsection{Baseline: Single-Component Dark Matter ($N=1$)}
\label{sec:results_single}

We begin by revisiting the minimal scenario in which the entire Galactic Center Excess (GCE) is attributed to the annihilation of a single dark-matter species. To establish a robust baseline for comparison, we adopt the comprehensive spectral analysis by Calore et al.~\cite{Calore:2014nla}, while employing the covariance matrix provided by \cite{Cholis:2021rpp}. As expected, when using data from the full $40^\circ\times40^\circ$ region around the Galactic Center, the fit is very poor, with no annihilation channel yielding an acceptable description of the data. However, restricting the analysis to the Southern sky, where contamination from diffuse Galactic emission is significantly reduced compared to the Northern region, leads to a substantially improved fit \cite{Cholis:2021rpp}. From this point onward, we therefore perform our analysis using only the Southern-sky data set.

Table~\ref{tab:single_component_calore} summarizes the best-fit parameters for a wide range of Standard Model final states. The results clearly demonstrate that the majority of theoretically motivated WIMP channels fail to reproduce the GCE spectrum when taken in isolation.

\begin{itemize}
    \item \textbf{Leptonic channels} ($\tau^+\tau^-$) yield spectra that are too hard and peaked, resulting in high $\chi^2$ values and negligible p-values ($p < 0.001$).
    \item \textbf{Heavy Quarks and Bosons} ($t\bar{t}, W^+W^-, ZZ$) are kinematically constrained to high masses ($m_\chi \gtrsim 80-175$ GeV). Consequently, their annihilation spectra peak at energies significantly higher than the observed excess, failing to match the characteristic turnover at $\sim 2$ GeV.
    \item \textbf{Higgs channel} ($hh$): While annihilation into Higgs bosons (which subsequently decay to $b\bar{b}$) provides a somewhat better fit than the gauge bosons, it remains statistically disfavored compared to direct $b\bar{b}$ production.
\end{itemize}

\begin{figure*}[h!]
  \centering
  \begin{minipage}[b]{0.55\textwidth}
    \includegraphics[width=\linewidth]{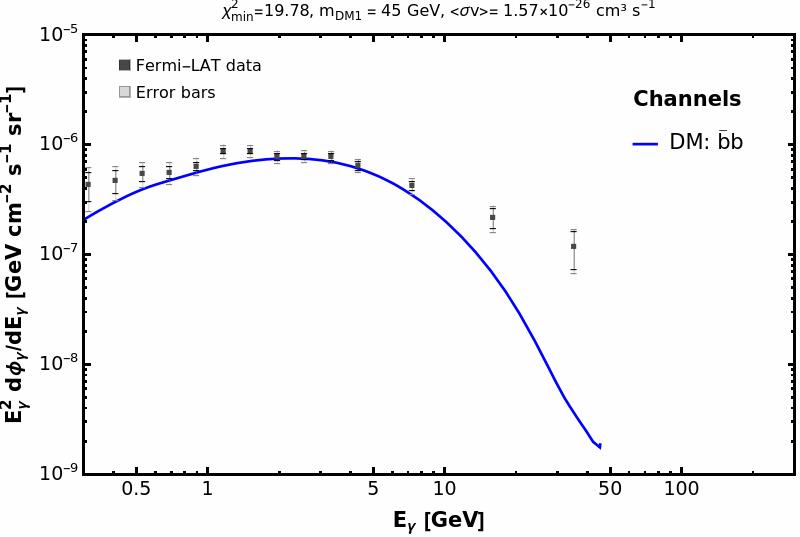}
    \label{fig:fig1}
  \end{minipage}
  \hfill
  \begin{minipage}[b]{0.4\textwidth}
    \includegraphics[width=\linewidth]{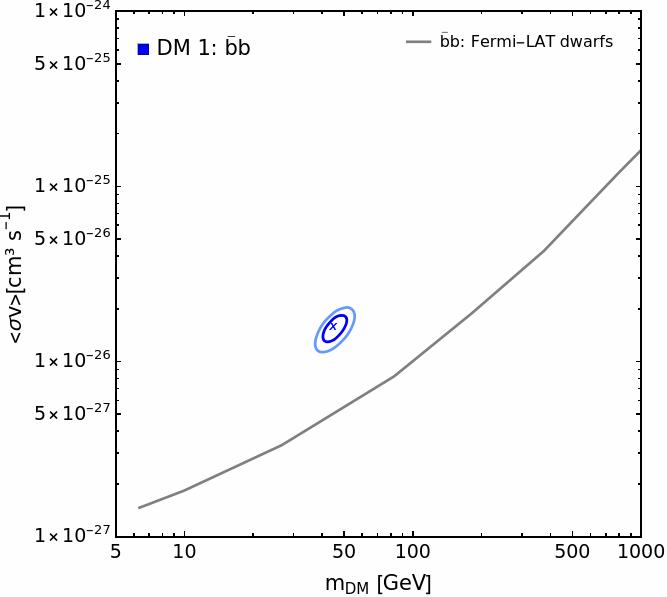}
    \label{fig:fig2}
  \end{minipage}
  \caption{Predicted flux for one DM component annihilating into $\bar{b}b$ compared with the Fermi-LAT data, including statistical and systematic uncertainties (left panel) and the benchmark point including $1\sigma$ and $2\sigma$ contours compared to the dSphs limits.}
  \label{fig:single}
\end{figure*}

Consistent with the consensus in the field, the annihilation into $b\bar{b}$ and light quarks remains the only single-channel solution that provides a statistically acceptable description of the data (with a p-value between 0.01 and 0.07). In Fig.~\ref{fig:single}, we show the predicted gamma-ray flux compared with the Fermi-LAT data, including the statistical and systematic uncertainties (left panel), and, in the right panel, the comparison with the combined limits from dwarf spheroidal galaxies (dSphs) \cite{Fermi-LAT:2025gei}. However, as we demonstrate in the following sections, the ``failed'' channels listed in Table~\ref{tab:single_component_calore}—particularly the heavy bosons and leptons—can play a critical role in multi-component scenarios.

\begin{table}[t!]
    \centering
    \renewcommand{\arraystretch}{1.1}
    \setlength{\tabcolsep}{6pt}
    \begin{tabular}{l c c c c}
    \hline
    \multicolumn{5}{c}{$40^\circ\times40^\circ$ sky map}\\ \hline
    \textbf{Channel} & \textbf{Mass $m_\chi$ [GeV]} & \textbf{$\langle \sigma v \rangle$ [$10^{-26}$ cm$^3$/s]} & \textbf{$\chi^2_{\text{min}}$} & \textbf{p-value} \\
    \hline
    \hline
    $b\bar b$    &    45    &    1.38    &    51.08    &    $8.9\times10^{-7}$    \\
    $\tau^+\tau^-$    &    10    &    0.28    &    53.85    &    $2.9\times10^{-7}$    \\  
    $q\bar q$    &    21    &    0.60    &    60.83    &    $1.6\times10^{-8}$    \\  
    $hh$    &    126    &    4.78    &    70.21    &    $2.92\times10^{-10}$    \\  
    $c\bar c$    &    28    &    1.00    &    92.07    &    $1.96\times10^{-14}$    \\  
    $ZZ$    &    91    &    3.59    &    93.66    &    $9.64\times10^{-15}$    \\  
    $W^+W^-$    &    80    &    2.83    &    102.01    &    $2.24\times10^{-16}$    \\  
    $t\bar t$    &    173    &    4.22    &    122.60    &    $1.87\times10^{-20}$    \\  
    \hline
    \multicolumn{5}{c}{Southern sky map}\\ \hline
    \textbf{Channel} & \textbf{Mass $m_\chi$ [GeV]} & \textbf{$\langle \sigma v \rangle$ [$10^{-26}$ cm$^3$/s]} & \textbf{$\chi^2_{\text{min}}$} & \textbf{p-value} \\
    \hline
    \hline
$b\bar b$    &    45    &    1.57    &    19.78    &    0.071    \\  
$q\bar q$    &    22    &    0.73    &    23.07    &    0.027    \\  
$c\bar c$    &    19    &    0.63    &    24.33    &    0.018    \\  
$hh$    &    126    &    4.70    &    38.02    &    0.00015    \\  
$\tau^+\tau^-$    &    10    &    0.32    &    38.51    &    0.00013    \\  
$ZZ$    &    91    &    3.66    &    53.32    &    $3.6\times10^{-7}$    \\
$W^+W^-$    &    80    &    3.00    &    58.68    &    $3.9\times10^{-8}$    \\
$t\bar t$    &    173    &    4.47    &    77.33    &    $1.3\times10^{-11}$    \\    \hline
    \end{tabular}
    \caption{Best-fit parameters for single-component dark matter ($N=1$) annihilating into various Standard Model final states. We consider the $40^\circ\times40^\circ$ and just the Southern region of the sky. We report the minimum $\chi^2$ and the associated p-value for 14 degrees of freedom. Only the Southern sky map for quarks provides a statistically acceptable fit; the other channels yield high $\chi^2$ values and are strongly disfavored as sole explanations for the Galactic Center Excess. See the text for details.}
    \label{tab:single_component_calore}
\end{table}

Finally, it is worth explicitly noting that the limitations of the single-component hypothesis cannot be resolved simply by allowing for mixed annihilation channels (e.g., a single particle annihilating into both $b\bar{b}$ and $\tau^+\tau^-$). As demonstrated in the detailed spectral analysis by Calore et al.~\cite{Calore:2014nla,Cholis:2021rpp}, relaxing the assumption of exclusive final states yields only marginal statistical improvement over the pure $b\bar{b}$ benchmark. The fundamental restriction remains the existence of a single dark matter mass scale $m_\chi$. This rigid kinematic constraint prevents a single particle from simultaneously populating the low-energy emission (which favors lighter masses) and the high-energy tail (which requires heavier masses) effectively, even when the branching ratios are treated as free parameters. Given that single-component models—whether exclusive or mixed—exhibit this persistent structural tension, we proceed to investigate the multicomponent framework, where distinct mass scales can naturally accommodate the detailed spectral morphology of the excess.

\subsection{Two-Component Dark Matter ($N=2$)}
\label{sec:results_two_component}

We now extend the analysis to a dark sector composed of two distinct species, $\chi_1$ and $\chi_2$. This framework introduces the necessary kinematic freedom to decouple the low-energy and high-energy spectral features of the GCE. We divide our analysis into two parts: first, considering fixed exclusive annihilation channels for each species, and second, allowing for general mixed branching ratios.

\setlength{\tabcolsep}{0pt}
\begin{table}[!ht]
\small
    \centering
    \begin{tabularx}{0.85\textwidth}{>{\centering\arraybackslash}X>{\centering\arraybackslash}X>{\centering\arraybackslash}X>{\centering\arraybackslash}X>{\centering\arraybackslash}X>{\centering\arraybackslash}X>{\centering\arraybackslash}X>{\centering\arraybackslash}X}
    \hline
        $\chi_1$ & $\chi_2$  & \multicolumn{2}{c}{$m_\chi$ (GeV)} & \multicolumn{2}{c}{$\langle \sigma v \rangle^{eff}$\footnote{$\times10^{-26}$ (cm$^3$s$^{-1}$)}}  & $\chi^2_{min}$ &  p-value \\ \hline
        \hline
    $Ch_1$ & $Ch_2$ & $\chi_1$ & $\chi_2$ & $\chi_1$ & $\chi_2$ & & \\
   \hline
\rowcolor{gray!10}   $b\bar b$    &    $W^+W^-$    &    34    &    192    &    1.25    &    3.24    &    9.14    &    0.52    \\  
$b\bar b$    &    $ZZ$    &    34    &    224    &    1.24    &    3.79    &    9.15    &    0.52    \\  
$b\bar b$    &    $q\bar q$    &    34    &    127    &    1.25    &    1.53    &    9.17    &    0.52    \\  
$b\bar b$    &    $hh$    &    34    &    371    &    1.27    &    5.02    &    9.18    &    0.52    \\  
$b\bar b$    &    $c\bar c$    &    34    &    113    &    1.24    &    1.34    &    9.18    &    0.52    \\  
$b\bar b$    &    $t\bar t$    &    34    &    363    &    1.27    &    5.21    &    9.18    &    0.52    \\  
$b\bar b$    &    $b\bar b$    &    287    &    34    &    3.32    &    1.30    &    9.24    &    0.51    \\  
$b\bar b$    &    $\tau^+\tau^-$    &    39    &    65    &    1.59    &    0.72    &    9.55    &    0.48    \\  
$q\bar q$    &    $ZZ$    &    15    &    187    &    0.50    &    4.01    &    9.67    &    0.47    \\  
\rowcolor{gray!10} $q\bar q$    &    $W^+W^-$    &    15    &    165    &    0.51    &    3.47    &    9.71    &    0.47    \\  
$q\bar q$    &    $hh$    &    15    &    302    &    0.51    &    5.17    &    9.74    &    0.46    \\  
$q\bar q$    &    $t\bar t$    &    15    &    308    &    0.52    &    5.45    &    9.77    &    0.46    \\  
$q\bar q$    &    $q\bar q$    &    16    &    119    &    0.54    &    1.63    &    9.80    &    0.46    \\  
$c\bar c$    &    $q\bar q$    &    113    &    16    &    1.43    &    0.56    &    9.82    &    0.46    \\  
$c\bar c$    &    $ZZ$    &    13    &    176    &    0.41    &    4.04    &    9.84    &    0.45    \\  
$c\bar c$    &    $W^+W^-$    &    14    &    164    &    0.44    &    3.46    &    9.92    &    0.45    \\  
$c\bar c$    &    $t\bar t$    &    13    &    279    &    0.42    &    5.30    &    9.95    &    0.44    \\  
$c\bar c$    &    $c\bar c$    &    14    &    103    &    0.46    &    1.43    &    10.00    &    0.44    \\
$c\bar c$    &    $hh$    &    11    &    205    &    0.31    &    4.88    &    10.22    &    0.42    \\  
$\tau^+\tau^-$    &    $hh$    &    5    &    136    &    0.11    &    4.50    &    13.21    &    0.21    \\  
\rowcolor{gray!10} $\tau^+\tau^-$    &    $ZZ$    &    6    &    91    &    0.13    &    3.28    &    14.44    &    0.15    \\     
\rowcolor{gray!10} $\tau^+\tau^-$    &    $W^+W^-$    &    6    &    85    &    0.15    &    2.80    &    14.84    &    0.14    \\ 
\hline
    \end{tabularx}
    \caption{Results considering two DM candidates taking the Southern region of the sky. We observe that the fit improves considerably, suggesting several possible explanations for the GeV excess.}
    \label{tab:two_component_exclusive}
\end{table}

We first investigate the simplified scenario where each dark matter component annihilates 100\% into a specific Standard Model final state. We perform a grid scan over all possible pairwise combinations of the channels listed in Table~\ref{tab:single_component_calore}. The best-fit parameters for the most successful combinations are summarized in Table~\ref{tab:two_component_exclusive}. We find that, as expected, the addition of a second component reduces the minimum $\chi^2$ from $19.78$ (for the single-component $b\bar{b}$ baseline) to $9.14$. 

To quantify the statistical significance of this improvement and penalize the addition of new parameters, we evaluate the difference in the Akaike Information Criterion, $\Delta \text{AIC}$. The baseline single-component model ($b\bar{b}$) features $k=2$ free parameters, establishing $\text{AIC}_{N=1} = 19.78 + 4 = 23.78$. Our best-fit exclusive two-component model ($b\bar{b} + W^+W^-$) introduces $k=4$ free parameters, yielding $\text{AIC}_{N=2} = 9.14 + 8 = 17.14$. The resulting difference is $\Delta \text{AIC} = -6.64$. According to the scale established in Section~\ref{sec:analysis}, this value represents substantial to strong statistical evidence in favor of the multicomponent scenario over the standard baseline.

\begin{figure*}[t!]
  \centering
  \begin{minipage}[b]{0.49\textwidth}
    \includegraphics[width=\linewidth]{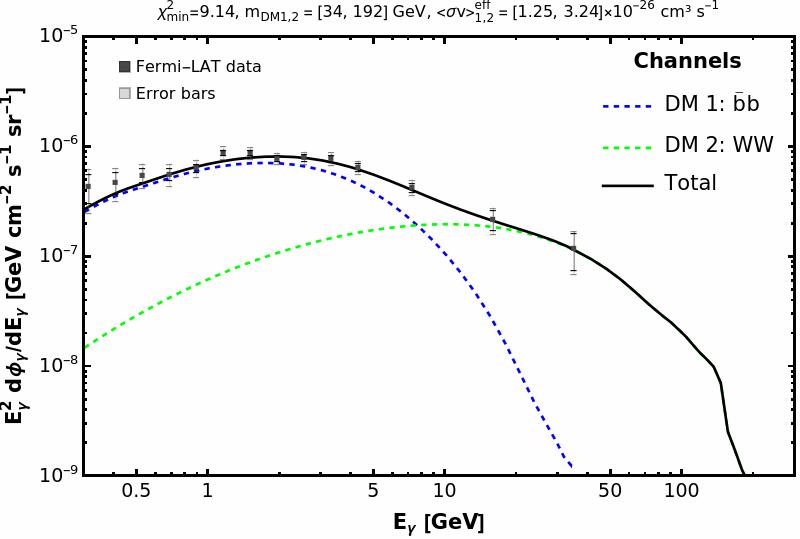}
  \end{minipage}
  \hfill
  \begin{minipage}[b]{0.49\textwidth}
    \includegraphics[width=\linewidth]{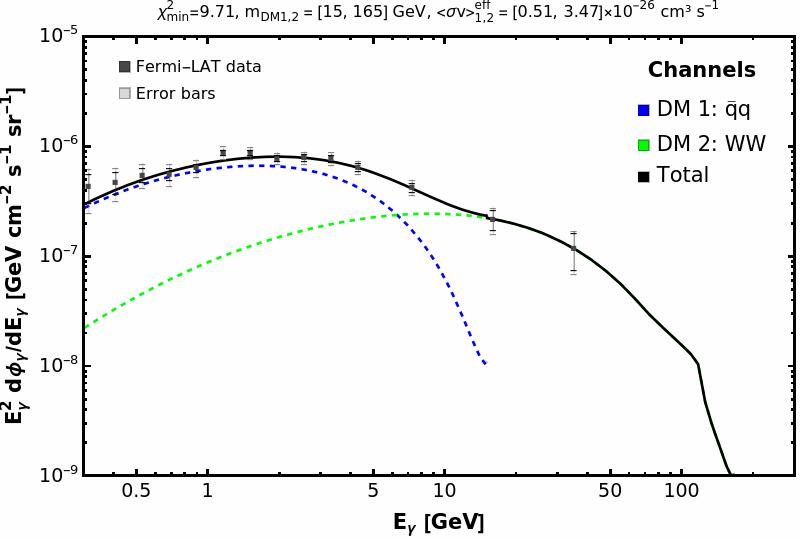}
  \end{minipage}
  \\[5mm]
  \begin{minipage}[b]{0.49\textwidth}
    \includegraphics[width=\linewidth]{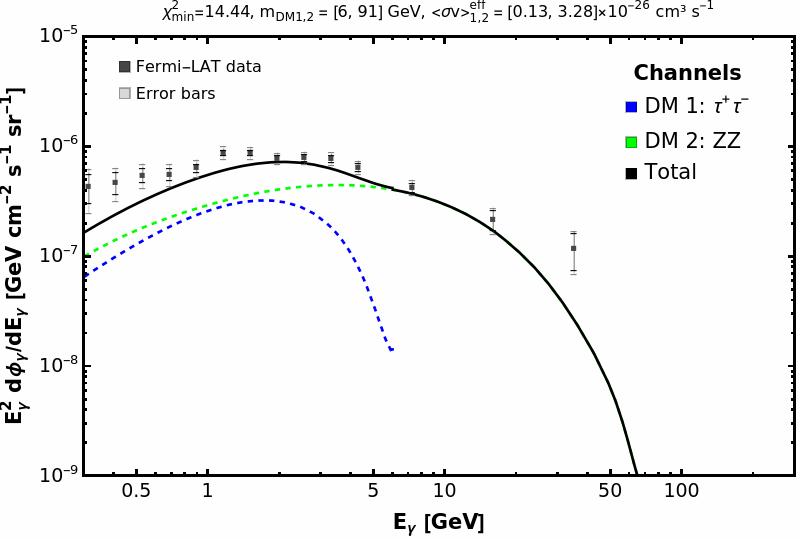}
  \end{minipage}
  \hfill
    \begin{minipage}[b]{0.49\textwidth}
    \includegraphics[width=\linewidth]{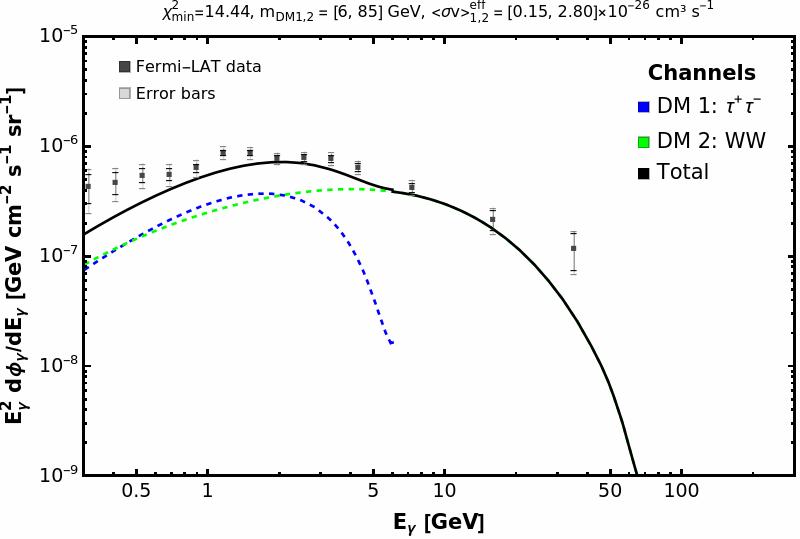}
  \end{minipage}
  \hfill
  \caption{Fluxes generated by two dark matter components combined, given the best fit to the Fermi-LAT data for several possible channels, $\bar{b}b$ and $WW$, $\bar{q}q$ and $WW$, $\tau^+\tau^-$ and $ZZ$, and $\tau^+\tau^-$ and $WW$, with the first contribution given in blue, the second one in green and the combined flux in black. The gray dots are the Fermi-LAT data, including statistical and systematic errors. }
  \label{fig:fits2DM}
\end{figure*}

Notice that most solutions naturally fall into a ``Heavy + Light'' hierarchy, where the lighter dark matter particle is responsible for fitting the peak of the spectrum while the heavier particle populates the high-energy tail. Two examples of this type are shown in the first row of Figure~\ref{fig:fits2DM}, corresponding to the final states $b\bar{b} + ZZ$ and $q\bar{q} + t\bar{t}$.

Another important result of our analysis is the fact that by combining final states that by themselves cannot explain the excess, we can get a reasonable fit to the data. Two examples of this type are $b\bar b + W^+W^-$ and $\tau^+\tau^- + ZZ$, as illustrated in Figure~\ref{fig:fits2DM}. This means that two-component scenarios can effectively rescue dark matter models with such final states as viable solutions to the GCE.

\subsubsection{Mixed Annihilation Channels}
\label{sec:mixed_channels}

For completeness, we relax the assumption of exclusive final states and allow the dark matter components to annihilate into multiple channels simultaneously. Given the large combinatorial space of possible Standard Model final states, we structure this analysis into two distinct scenarios of increasing complexity:

\begin{enumerate}
    \item \textbf{Single-Mixed Scenario (5 Parameters):} In this configuration, we assume an asymmetric setup where one dark matter component annihilates into a mixture of two final states (governed by a free branching fraction), while the second component remains exclusive to a single channel. This introduces five free parameters: the two masses, the two effective cross-sections, and one branching fraction.
    \item \textbf{Double-Mixed Scenario (6 Parameters):} In this configuration, we allow both dark matter components to annihilate into mixtures of two different final states. This introduces six free parameters: the two masses, the two effective cross-sections, and two independent branching fractions.
\end{enumerate}

Since the number of possible channel combinations is extremely large, we perform a broad scan and report only the most significant results. Tables~\ref{tab:mixed_scenario_1} and \ref{tab:mixed_scenario_2} list the top solutions with the lowest minimum $\chi^2$ for the single-mixed and double-mixed scenarios, respectively.

Two key conclusions emerge from this general analysis:
\begin{enumerate}
    \item \textbf{Persistence of the Mass Hierarchy:} Even with the freedom to adjust branching ratios, the favored regions of parameter space largely reproduce the phenomenological pattern found in the exclusive analysis. The fits consistently prefer one lighter component (to fit the peak) and one heavier component (to fit the tail).
    \item \textbf{Saturation of the Fit:} The minimum $\chi^2$ values obtained in these mixed scenarios (e.g., $8.97$ and $8.93$) are remarkably similar to those found in the exclusive analysis (Table~\ref{tab:two_component_exclusive}). Because these minor improvements in $\chi^2$ come at the cost of additional free parameters ($k=5$ and $k=6$, respectively), they are heavily penalized by the Akaike Information Criterion. The resulting $\Delta \text{AIC}$ values for the mixed models ($-4.81$ and $-2.85$) demonstrate weaker statistical preference compared to the optimal exclusive model ($\Delta \text{AIC} = -6.64$). This mathematically confirms that the simple superposition of two exclusive spectra is sufficient to capture the morphology of the GCE, and the "Heavy + Light" mass hierarchy is the primary driver of the successful fit.
\end{enumerate}

\begin{table*}[h!]
    \centering
    \begin{tabularx}{0.85\textwidth}{>{\centering\arraybackslash}X>
    {\centering\arraybackslash}X>
    {\centering\arraybackslash}X>
    {\centering\arraybackslash}X>{\centering\arraybackslash}X>{\centering\arraybackslash}X>{\centering\arraybackslash}X>{\centering\arraybackslash}X>{\centering\arraybackslash}X>{\centering\arraybackslash}X>{\centering\arraybackslash}X}
    \hline
        \multicolumn{2}{c}{$\chi_1$} & $\chi_2$  & $BR_1$  & \multicolumn{2}{c}{$m_\chi$ (GeV)} & \multicolumn{2}{c}{$\langle \sigma v \rangle^{eff}$\footnote{All the effective cross-sections are given in terms of $10^{-26}$} (cm$^3$s$^{-1})$ } & $ \chi^2_{min}$ &  p-value \\ \hline
        \hline
    $\ \ Ch_1 \ \ $ & $Ch_2$ & $Ch_3$  & & $\chi_1$ & $\chi_2$ & $\chi_1$ & $\chi_2$ & \\
    \hline
$b\bar b$	&	$hh$	&	$WW$	&	0.84	&	39	&	226	&	1.62	&	3.25	&	8.97	&	0.44	\\
$\tau^+\tau^-$	&	$\tau^+\tau^-$	&	$t\bar t$	&	0.97	&	72	&	89	&	0.71	&	4.41	&	8.98	&	0.44	\\
$q\bar q$	&	$\tau^+\tau^-$	&	$t\bar t$	&	0.25	&	74	&	85	&	0.85	&	4.11	&	8.99	&	0.44	\\
$\tau^+\tau^-$	&	$t\bar t$	&	$W^+W^-$	&	0.10	&	84	&	173	&	4.38	&	1.45	&	9.03	&	0.43	\\
$b\bar b$	&	$b\bar b$	&	$t\bar t$	&	1.00	&	381	&	83	&	3.59	&	3.95	&	9.03	&	0.43	\\
$b\bar b$	&	$c\bar c$	&	$t\bar t$	&	1.00	&	381	&	83	&	3.59	&	3.95	&	9.03	&	0.43	\\
$b\bar b$	&	$q\bar q$	&	$t\bar t$	&	1.00	&	381	&	83	&	3.59	&	3.95	&	9.03	&	0.43	\\
$b\bar b$	&	$\tau^+\tau^-$	&	$t\bar t$	&	1.00	&	381	&	83	&	3.59	&	3.95	&	9.03	&	0.43	\\
$\tau^+\tau^-$	&	$t\bar t$	&	$t\bar t$	&	0.10	&	84	&	312	&	4.40	&	2.20	&	9.04	&	0.43	\\
$b\bar b$	&	$hh$	&	$t\bar t$	&	0.83	&	39	&	394	&	1.62	&	4.94	&	9.05	&	0.43	\\

\hline
    \end{tabularx}
    \caption{Results considering two DM candidates with one of them annihilating in two different channels with branching ratio given by $BR_1$.}
    \label{tab:mixed_scenario_1}
\end{table*}

\begin{table*}[h!]
    \centering
    \begin{tabularx}{0.95\textwidth}{>{\centering\arraybackslash}X>
    {\centering\arraybackslash}X>
    {\centering\arraybackslash}X>
    {\centering\arraybackslash}X>{\centering\arraybackslash}X>{\centering\arraybackslash}X>{\centering\arraybackslash}X>{\centering\arraybackslash}X>{\centering\arraybackslash}X>{\centering\arraybackslash}X>{\centering\arraybackslash}X>{\centering\arraybackslash}X}
    \hline
        \multicolumn{2}{c}{$\chi_1$} & \multicolumn{2}{c}{$\chi_2$} & $BR_1$ & $BR_2$ & \multicolumn{2}{c}{$m_\chi$ (GeV)} & \multicolumn{2}{c}{$\langle \sigma v \rangle^{eff}$\footnote{All the effective cross-sections are given in terms of $10^{-26}$} (cm$^3$s$^{-1})$ } & $ \chi^2_{min}$ &  p-value \\ \hline
        \hline
    $\ \ Ch_1 \ \ $ & $Ch_2$ & $Ch_3$& $Ch_4$  &  & & $\chi_1$ & $\chi_2$ & $\chi_1$ & $\chi_2$ & \\
    \hline
$b\bar b$	&	$b\bar b$	&	$\tau^+\tau^-$	&	$t\bar t$	&	0.40	&	0.15	&	119	&	82	&	0.68	&	4.44	&	8.93	&	0.35	\\
$b\bar b$	&	$\tau^+\tau^-$	&	$\tau^+\tau^-$	&	$t\bar t$	&	1.00	&	0.15	&	119	&	82	&	0.68	&	4.44	&	8.93	&	0.35	\\
$q\bar q$	&	$\tau^+\tau^-$	&	$\tau^+\tau^-$	&	$t\bar t$	&	0.10	&	0.14	&	36	&	85	&	0.16	&	4.88	&	8.95	&	0.35	\\
$c\bar c$	&	$\tau^+\tau^-$	&	$\tau^+\tau^-$	&	$t\bar t$	&	0.42	&	0.11	&	58	&	85	&	0.34	&	4.62	&	8.98	&	0.34	\\
$\tau^+\tau^-$	&	$\tau^+\tau^-$	&	$\tau^+\tau^-$	&	$t\bar t$	&	0.55	&	0.10	&	58	&	89	&	0.27	&	4.89	&	8.99	&	0.34	\\
$b\bar b$	&	$b\bar b$	&	$c\bar c$	&	$hh$	&	0.22	&	0.22	&	21	&	83	&	0.17	&	3.75	&	8.99	&	0.34	\\
$c\bar c$	&	$c\bar c$	&	$\tau^+\tau^-$	&	$t\bar t$	&	0.88	&	0.12	&	82	&	84	&	0.41	&	4.51	&	9.00	&	0.34	\\
$q\bar q$	&	$q\bar q$	&	$\tau^+\tau^-$	&	$t\bar t$	&	0.11	&	0.10	&	108	&	84	&	0.61	&	4.42	&	9.01	&	0.34	\\
$c\bar c$	&	$q\bar q$	&	$\tau^+\tau^-$	&	$t\bar t$	&	0.10	&	0.10	&	117	&	84	&	0.66	&	4.43	&	9.01	&	0.34	\\
$b\bar b$	&	$b\bar b$	&	$q\bar q$	&	$hh$	&	0.40	&	0.24	&	21	&	82	&	0.17	&	3.69	&	9.02	&	0.34	\\

\hline
    \end{tabularx}
    \caption{Results considering two DM candidates, with two of them annihilating in two different channels with branching ratios given by $BR_1$ and $BR_2$.}
    \label{tab:mixed_scenario_2}
\end{table*}

\subsection{Three-Component Dark Matter ($N=3$)}
\label{sec:results_three_component}

In the previous section, we observed that increasing the model complexity by allowing for mixed branching ratios yields only marginal improvements in the goodness-of-fit. This suggests that the limitation of the two-component model is not the composition of the final states, but perhaps the rigidity of the kinematic edges defined by the two mass scales. 

Motivated by this, we extend our framework to a three-component dark sector ($N=3$). This introduces a third independent mass scale, potentially allowing the model to populate the spectral features of the GCE with even greater precision. We analyze two specific scenarios of increasing complexity.

\subsubsection{Exclusive Annihilation Channels}
\label{sec:three_component_exclusive}

First, we consider the case where all three dark matter components annihilate exclusively into single Standard Model final states. This configuration introduces six free parameters: three masses ($m_{\chi,1}, m_{\chi,2}, m_{\chi,3}$) and three effective cross-sections.

We perform a scan over the relevant channel combinations. The most statistically significant solutions are summarized in Table~\ref{tab:three_component_exclusive}. Interestingly, we observe a recurrent phenomenological pattern in the best-fit regions, characterized by a distinct mass hierarchy with one light, one intermediate, and one heavy component.

However, the most striking result of this expansion is that the minimum $\chi^2$ values obtained for $N=3$ (e.g., $8.91$) are essentially identical to those found for the two-component scenarios ($9.14$). This indicates a definitive ``saturation'' of the fit. Despite the addition of two extra degrees of freedom, the model is unable to further reduce the test statistic in any meaningful way. This suggests that the two-component framework already captures the full physical information contained in the GCE spectral shape, and the third component merely attempts to fit statistical noise or minor systematic fluctuations without adding explanatory power. 

This lack of improvement is heavily penalized when comparing the model to the single-component baseline. Evaluating the best exclusive $N=3$ model ($k=6$) against the single-component $b\bar{b}$ baseline ($\text{AIC}_{N=1} = 23.78$) yields $\Delta\text{AIC} = (8.91 + 12) - 23.78 = -2.87$. While this remains a better fit than the $N=1$ case, the fact that the $\Delta\text{AIC}$ is significantly less negative than the $N=2$ result ($-6.64$) provides strong evidence that $N=2$ is the optimal dark sector complexity for explaining the Fermi-LAT data.

\begin{table}[t!]
    \centering
    \begin{tabularx}{0.95\textwidth}{>{\centering\arraybackslash}X>{\centering\arraybackslash}X>{\centering\arraybackslash}X>{\centering\arraybackslash}X>{\centering\arraybackslash}X>{\centering\arraybackslash}X>{\centering\arraybackslash}X>{\centering\arraybackslash}X>{\centering\arraybackslash}X>{\centering\arraybackslash}X>{\centering\arraybackslash}X}
    \hline
    $\chi_1$ & $\chi_2$  & $\chi_3$ & \multicolumn{3}{c}{$m_\chi$ (GeV)} & \multicolumn{3}{c}{$\langle \sigma v \rangle^{eff}$\footnote{All the effective cross-sections are given in terms of $10^{-26}$ (cm$^3$s$^{-1})$}} & $\chi^2_{min}$   & p-value \\ \hline
    \hline
    $Ch_1$ & $Ch_2$ & $Ch_3$ & $\chi_1$ & $\chi_2$ & $\chi_3$ & $\chi_1$ & $\chi_2$ & $\chi_3$ & & \\
    \hline
$b\bar b$    &    $b\bar b$    &    $c\bar c$    &    32    &    118    &    389    &    1.07    &    1.26    &    2.35    &    8.91    &    0.35    \\
$b\bar b$    &    $b\bar b$    &    $\tau^+\tau^-$    &    30    &    126    &    464    &    0.98    &    1.97    &    6.86    &    8.94    &    0.35    \\
$b\bar b$    &    $t\bar t$    &    $t\bar t$    &    34    &    218    &    676    &    1.21    &    1.97    &    4.74    &    9.09    &    0.33    \\
$b\bar b$    &    $hh$    &    $t\bar t$    &    34    &    777    &    234    &    1.23    &    4.29    &    2.34    &    9.10    &    0.33    \\
$b\bar b$    &    $q\bar q$    &    $t\bar t$    &    34    &    272    &    234    &    1.21    &    1.38    &    2.29    &    9.10    &    0.33    \\
$b\bar b$    &    $\tau^+\tau^-$    &    $\tau^+\tau^-$    &    35    &    35    &    339    &    1.40    &    0.41    &    4.37    &    9.12    &    0.33    \\
$b\bar b$    &    $ZZ$    &    $t\bar t$    &    34    &    180    &    1000    &    1.23    &    2.60    &    3.16    &    9.12    &    0.33    \\
$b\bar b$    &    $ZZ$    &    $W^+W^-$    &    34    &    180    &    916    &    1.22    &    2.79    &    2.90    &    9.12    &    0.33    \\
$b\bar b$    &    $q\bar q$    &    $hh$    &    34    &    180    &    204    &    1.21    &    1.33    &    1.31    &    9.12    &    0.33    \\
$b\bar b$    &    $c\bar c$    &    $\tau^+\tau^-$    &    34    &    83    &    419    &    1.21    &    0.94    &    3.32    &    9.12    &    0.33    \\
\hline
\end{tabularx}
    \caption{The 10 best results considering three DM candidates, each one annihilating 100\% into a single channel for the Southern sky region. As we can see, although they provide a good fit to the data, the inclusion of new degrees of freedom naturally degrades the p-value, as expected.}
    \label{tab:three_component_exclusive}
\end{table}

\subsubsection{Single-Mixed Scenario}
\label{sec:three_component_mixed}

Second, we consider a slightly more general scenario where one of the three dark matter particles is allowed to annihilate into a mixture of two final states, while the other two remain exclusive. This introduces a total of seven free parameters: 3 masses, 3 cross-sections, and 1 branching ratio.

The results for the most relevant channel combinations are displayed in Table~\ref{tab:three_component_mixed}. Consistent with our previous findings, the inclusion of branching ratio freedom does not yield a substantial reduction in $\chi^2$ relative to the exclusive $N=3$ case, nor relative to the optimal $N=2$ solutions. Here, the minimum $\chi^2 = 8.96$ requires $k=7$ parameters, yielding an $\text{AIC} = 22.96$. When compared to the single-component baseline ($\text{AIC}_{N=1} = 23.78$), the improvement is a negligible $\Delta\text{AIC} = -0.82$. This reinforces the conclusion that while increasing the complexity of the dark sector can technically accommodate the GCE data, the most statistically favored and efficient explanation remains the two-scale exclusive model.

\begin{table}[t!]
    \centering
    \begin{tabularx}{\textwidth}{>{\centering\arraybackslash}X>{\centering\arraybackslash}X>{\centering\arraybackslash}X>{\centering\arraybackslash}X>{\centering\arraybackslash}X>{\centering\arraybackslash}X>{\centering\arraybackslash}X>{\centering\arraybackslash}X>{\centering\arraybackslash}X>{\centering\arraybackslash}X>{\centering\arraybackslash}X>{\centering\arraybackslash}X>{\centering\arraybackslash}X}
    \hline
     \multicolumn{2}{c}{$\chi_1$} & $\chi_2$  & $\chi_3$ & & \multicolumn{3}{c}{$m_\chi$ (GeV)} & \multicolumn{3}{c}{$\langle \sigma v \rangle^{eff}$\footnote{All the effective cross-sections are given in terms of $10^{-26}$ (cm$^3$s$^{-1})$}} & $\chi^2_{min}$   & p-value \\ \hline
    \hline
    $Ch_1$ & $Ch_2$ & $Ch_3$ & $Ch_4$ & $BR_1$ & $\chi_1$ & $\chi_2$ & $\chi_3$ & $\chi_1$ & $\chi_2$ & $\chi_3$ & & \\
    \hline
$b\bar b$    &    $b\bar b$    &    $b\bar b$    &    $ZZ$    &    0.90    &    104    &    30    &    678    &    1.35    &    0.98    &    5.74    &    8.96    &    0.26    \\   
$b\bar b$    &    $b\bar b$    &    $b\bar b$    &    $c\bar c$    &    0.88    &    129    &    34    &    312    &    1.00    &    1.18    &    1.96    &    9.03    &    0.25    \\   
$b\bar b$    &    $\tau^+\tau^-$    &    $\tau^+\tau^-$    &    $t\bar t$    &    0.83    &    34    &    234    &    215    &    1.58    &    2.10    &    1.04    &    9.05    &    0.25    \\   
$b\bar b$    &    $b\bar b$    &    $hh$    &    $t\bar t$    &    0.10    &    32    &    1000    &    206    &    1.12    &    5.34    &    2.71    &    9.05    &    0.25    \\   
$b\bar b$    &    $b\bar b$    &    $hh$    &    $hh$    &    0.58    &    32    &    170    &    739    &    1.04    &    2.28    &    4.62    &    9.07    &    0.25    \\   
$b\bar b$    &    $\tau^+\tau^-$    &    $hh$    &    $W^+W^-$    &    0.90    &    32    &    170    &    901    &    1.26    &    1.64    &    7.25    &    9.08    &    0.25    \\   
$b\bar b$    &    $\tau^+\tau^-$    &    $\tau^+\tau^-$    &    $ZZ$    &    0.77    &    34    &    464    &    637    &    1.80    &    10.00    &    0.01    &    9.09    &    0.25    \\   
$b\bar b$    &    $\tau^+\tau^-$    &    $\tau^+\tau^-$    &    $\tau^+\tau^-$    &    0.77    &    34    &    500    &    385    &    1.80    &    6.34    &    3.04    &    9.09    &    0.25    \\   
$b\bar b$    &    $c\bar c$    &    $q\bar q$    &    $\tau^+\tau^-$    &    0.85    &    34    &    180    &    5    &    1.30    &    1.69    &    0.01    &    9.10    &    0.25    \\   
$b\bar b$    &    $b\bar b$    &    $t\bar t$    &    $t\bar t$    &    0.41    &    34    &    206    &    513    &    1.23    &    1.31    &    4.71    &    9.11    &    0.25    \\   
\hline
\end{tabularx}
    \caption{Results considering three DM candidates, with one of them annihilating into two final states.}
    \label{tab:three_component_mixed}
\end{table}

\section{Discussion}
\label{sec:discussion}

Our analysis demonstrates that while increasing model complexity naturally improves the goodness-of-fit, the Galactic Center Excess is most efficiently described by a two-component dark sector. As established by the AIC preferences in the previous sections, the transition from $N=1$ to $N=2$ provides a statistically robust improvement that justifies the additional parameters. Conversely, further expansion to $N=3$ or the introduction of mixed branching ratios results in overparameterization and diminishing returns; the marginal reduction in $\chi^2$ fails to outweigh the penalty of the increased degrees of freedom.

\subsection{Degeneracy and the Window of Viability}
\label{sec:viability_window}

A primary finding of this work is that the preference for $N=2$ dark matter is not restricted to a single, unique channel combination. Instead, the data identifies a broad \textit{window of viability}. By comparing the exclusive $N=2$ models to the $N=1$ baseline ($b\bar{b}$, $\chi^2 \approx 19.78$), we account for the introduction of two additional parameters ($\Delta k = 2$): a second mass scale and a second effective cross-section. 

Using the AIC-derived thresholds, we classify the phenomenological landscape of multi-component solutions:
\begin{itemize}
    \item \textbf{Strong-Preference Solutions ($\chi^2 < 10$):} These models yield $\Delta \text{AIC} < -6$ relative to the baseline, indicating a highly significant statistical improvement. A notable result of our scan is the abundance of solutions within this category; as shown in Table~\ref{tab:two_component_exclusive}, a wide variety of combinations—ranging from $b\bar{b} + ZZ$ and $b\bar{b} + t\bar{t}$ to $c\bar{c} + b\bar{b}$—all satisfy this stringent criterion. This high density of best-fit candidates suggests that the two-component framework identifies a robust phenomenological region where the spectral tension of the GCE is consistently resolved.
    
    \item \textbf{Positive-Preference Solutions ($10 < \chi^2 < 14$):} These models yield $-6 < \Delta \text{AIC} < -2$. A remarkable example in this category is the combination of $\tau^+\tau^-$ and $hh$. Neither final state provides an acceptable fit to the excess individually; however, they become statistically viable when combined within the $N=2$ framework.
    
    \item \textbf{Weak-Preference Solutions ($14 < \chi^2 < 16$):} These combinations provide a better fit than the $b\bar{b}$ baseline ($\Delta \text{AIC} < 0$) but are penalized by their complexity, offering only marginal statistical gains. A characteristic example is the pairing of $\tau^+\tau^-$ and $W^+W^-$ ($\chi^2 = 14.84$). Although the two-component framework improves the fit relative to the $N=1$ case, the inherently hard spectral features of these specific channels remain in slight tension with the broader, softer morphology of the observed excess.
\end{itemize}

This classification reveals a definitive ``resurrection'' of final states. In a standard $N=1$ framework, channels like $\tau^+\tau^-$, $hh$, and $t\bar{t}$ are typically excluded because their individual spectra cannot simultaneously match the GCE peak and high-energy tail. Within the $N=2$ framework, the second mass scale provides the necessary kinematic flexibility to anchor the $1-3$ GeV peak while independently populating the tail. This synergy suggests that the GCE may be masking a diverse dark sector where multiple mass scales cooperate to produce the observed spectral morphology.

\subsection{Constraints from Independent Searches}
\label{sec:external_constraints}

To validate the physical viability of the two-component scenarios, we contrast our best-fit regions against independent limits from Fermi-LAT Dwarf Spheroidal galaxies (dSphs) \cite{Fermi-LAT:2025gei} and AMS-02 cosmic-ray antiprotons \cite{Calore:2022stf}.

It is important to emphasize that these external bounds are subject to significant astrophysical uncertainties. For dSphs, the limits are sensitive to $J$-factor estimations derived from stellar kinematics, which can vary by factors of a few depending on the assumed dark matter halo profile \cite{Bonnivard:2015xpq,Ullio:2016kvy,Horigome:2022gge,10.1093/mnras/stag279}. For antiprotons, the constraints are highly dependent on the choice of cosmic-ray propagation models and solar modulation potentials \cite{Reinert:2017aga,Heisig:2020nse, DeLaTorreLuque:2024htu}. Consequently, these bounds serve as illustrative benchmarks rather than absolute exclusion criteria.

\begin{figure*}[t!]
  \centering
  \begin{minipage}[b]{0.48\textwidth}
    \includegraphics[width=\linewidth]{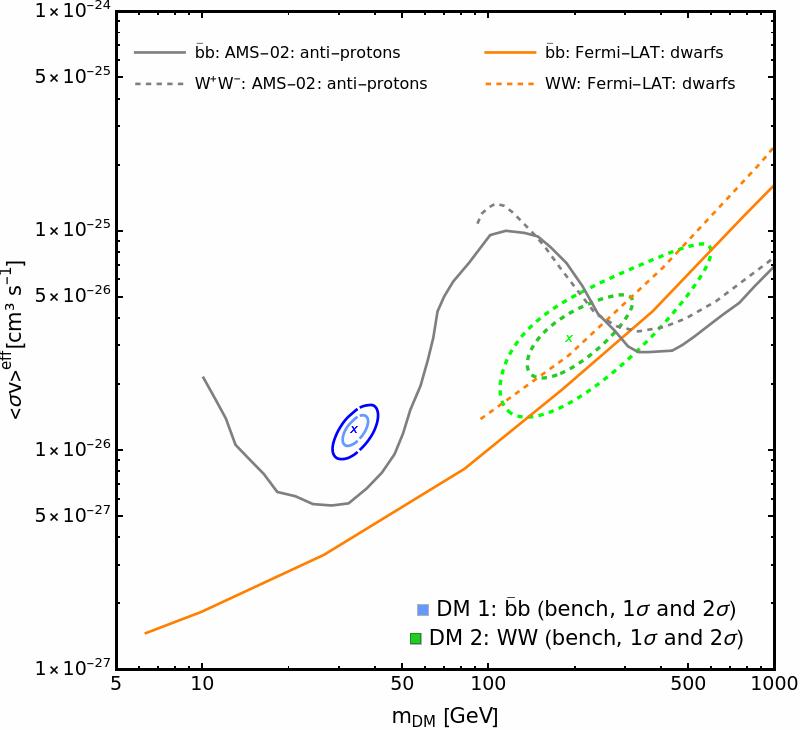}
    \label{fig:bw_comparison}
  \end{minipage}
  \hfill
  \begin{minipage}[b]{0.48\textwidth}
    \includegraphics[width=\linewidth]{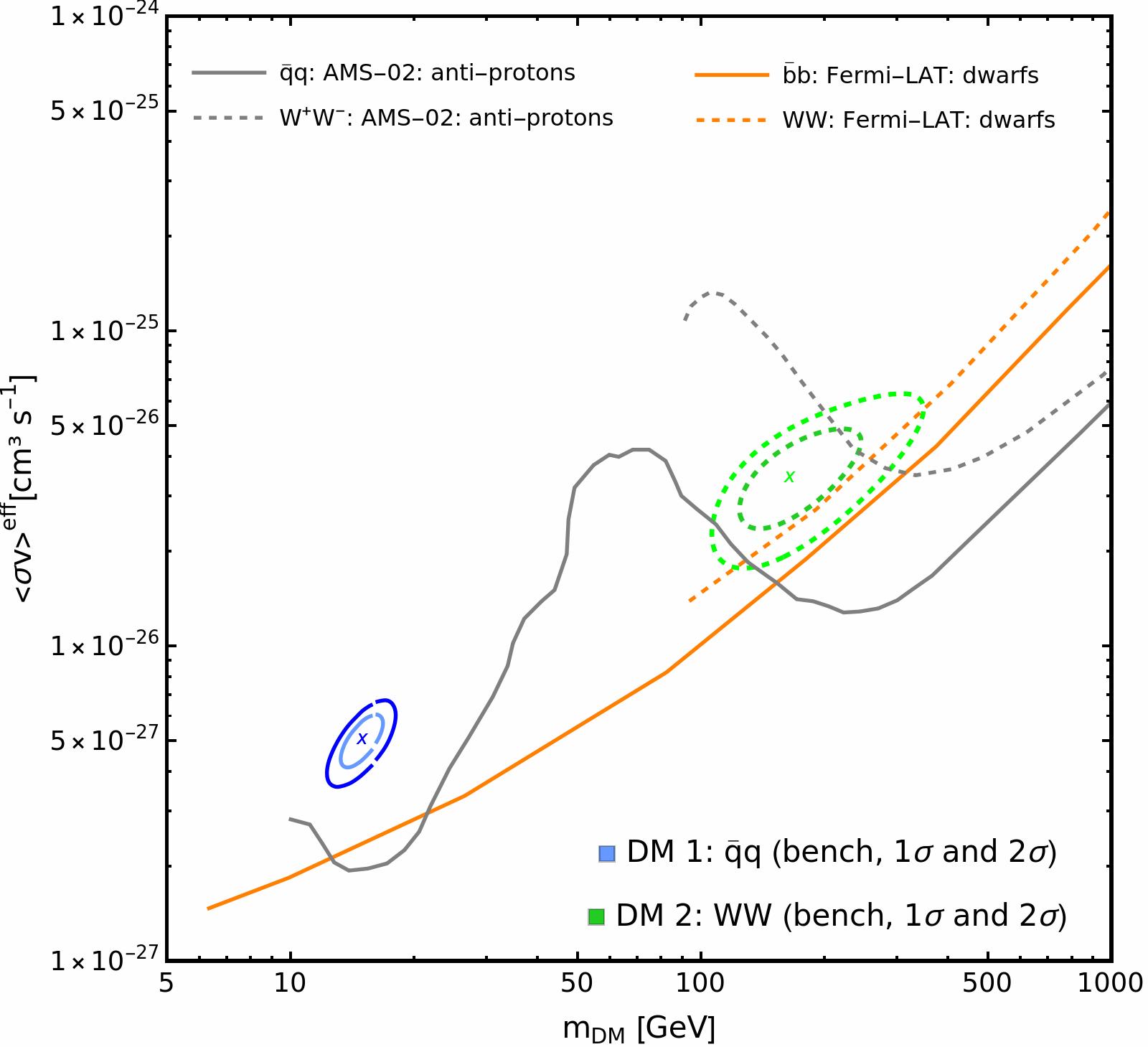}
    \label{fig:qw_comparison}
  \end{minipage}
  \\
  \begin{minipage}[b]{0.48\textwidth}
    \includegraphics[width=\linewidth]{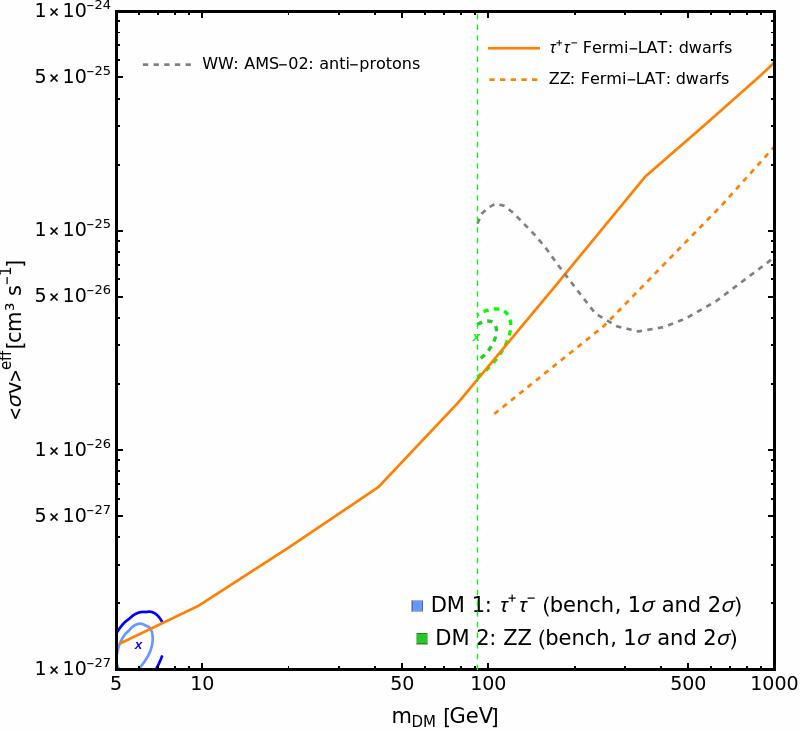}
    \label{fig:tauz_comparison}
  \end{minipage}
  \hfill
    \begin{minipage}[b]{0.48\textwidth}
    \includegraphics[width=\linewidth]{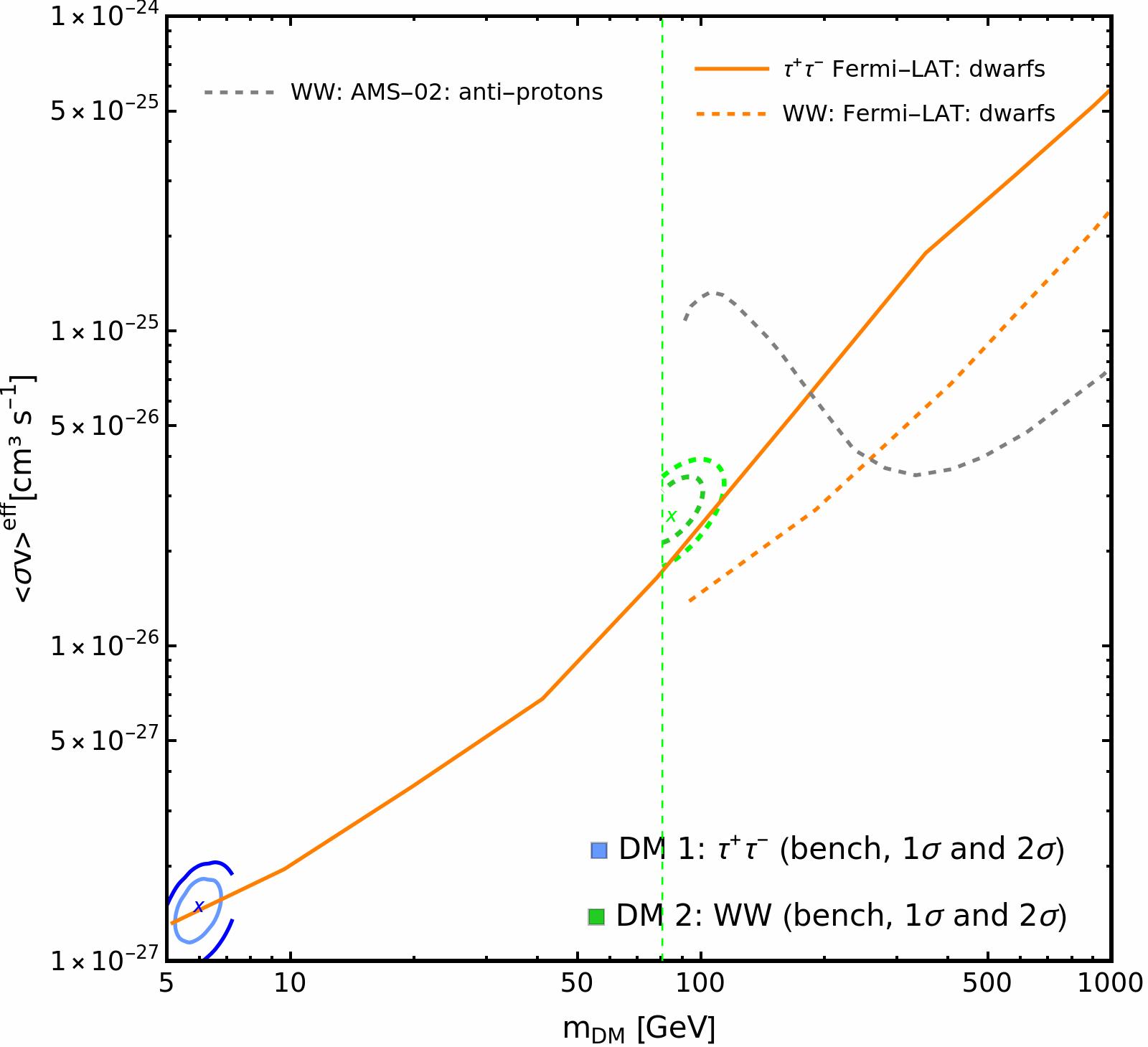}
    \label{fig:tauw_comparison}
  \end{minipage}
  \caption{The best-fit $1\sigma$ and $2\sigma$ regions for four representative scenarios in our exclusive two-component fit ($N=2$), compared with exclusion limits from Fermi-LAT dSphs and AMS-02 antiprotons. Systematic uncertainties in the $J$-factors and propagation models may shift these exclusion boundaries, potentially resolving regions of apparent tension.}
  \label{fig:comparison}
\end{figure*}

As shown in Figure~\ref{fig:comparison}, several of our $N=2$ best-fit regions—particularly those involving lepton pairs or heavy bosons—lie in regions of parameter space that remain broadly consistent with current null results when accounting for the inherent astrophysical systematic uncertainties \cite{Ando:2020yyk}. This alignment suggests that multi-component dark matter not only resolves the internal spectral tension of the GCE but also remains a globally viable candidate in the context of the wider indirect detection landscape.

\section{Conclusions}
\label{sec:conclusions}

In this work, we have provided a systematic analysis of the Galactic Center Excess (GCE) through the lens of multi-component dark matter. Motivated by the possibility that the dark sector may be non-minimal—rather than consisting of a single stable species—we investigate the phenomenological viability of scenarios containing two ($N=2$) and three ($N=3$) distinct particle species. Using the standard single-component hypothesis ($N=1$) as a statistical baseline, we rigorously evaluate whether the spectral morphology of the excess justifies a departure from the minimal WIMP paradigm in favor of a slightly expanded dark sector.

By employing the Akaike Information Criterion (AIC) to strictly balance fit quality against model complexity, we identify the exclusive two-component scenario ($N=2$) as the statistically preferred framework for the GCE. This model achieves a substantial improvement in fit quality relative to the $N=1$ baseline, yielding strong statistical evidence ($\Delta \text{AIC} \approx -6.6$) that the excess is better described by two distinct mass scales than by a single particle species. Furthermore, our analysis demonstrates that increasing the model complexity beyond this point yields diminishing returns; scenarios involving mixed branching ratios or three distinct components ($N=3$) fail to justify their additional degrees of freedom, resulting in a \emph{saturation} of the fit where the statistical support is weaker than in the exclusive $N=2$ case.

A central phenomenological consequence of this preference is the opening of a window of viability for particle physics models that were previously disfavored. We observe a significant degeneracy within the two-component parameter space, where a wide variety of channel combinations satisfy the criteria for a statistically preferred fit ($\chi^2 < 16$). Despite this variety, a clear morphological pattern emerges: the best-fit models consistently feature a light+heavy mass hierarchy. In this configuration, a lower-mass component anchors the spectral peak while a heavier species populates the high-energy tail. This synergy enables the statistical resurrection of annihilation final states—such as $t\bar{t}$, $ZZ$, and $hh$—that are typically ruled out in single-component analyses due to their high-mass thresholds.

Finally, we find that this multi-component framework offers a robust pathway to resolve potential tensions with independent indirect detection constraints. When viewed in light of the significant astrophysical uncertainties associated with $J$-factors in dwarf spheroidal galaxies and cosmic-ray propagation models for antiprotons, our results indicate that two-component scenarios can remain compatible with current null results. Ultimately, our results establish the two-component dark sector not merely as a viable alternative, but as the superior statistical and phenomenological framework for decoding the true nature of the Galactic Center Excess.

\begin{acknowledgments}
CNPq supports CS through grant numbers 304944/2025-4 and 406718/2025-3. CS and FSQ is supported by the São Paulo Research Foundation (FAPESP) through grant number 2021/01089-1. FSQ acknowledges support from Simons Foundation (Award Number:1023171-RC), CNPq 403521/2024-6, 408295/2021-0, 403521/2024-6, 406919/2025-9, 351851/2025-9, the FAPESP 2023/01197-4, ICTP-SAIFR 2021/14335-0, and the ANID-Millennium Science Initiative Program ICN2019\_044. This work is partially funded by FINEP under project 213/2024.\\

\end{acknowledgments}
\appendix

\bibliographystyle{JHEP}
\bibliography{references}

\end{document}